# Detection of gravitational waves in Michelson interferometer by the use of second order correlation functions

Y.Ben-Aryeh

Physics department, Technion-Israel Institute of Technology, Haifa 32000, Israel

e-mail:  phr65yb@physics.technion.ac.il

**Abstract**

The possibility of measuring the second order correlation function of the gravitational waves detectors' currents or photonumbers , and the observation of the gravitational signals by using a spectrum analyzer is discussed. The method is based on complicated data processing and is expected to be efficient for coherent periodic gravitational waves. It is suggested as an alternative method to the conventional one which is used now in the gravitational waves observatories.

## 1. Introduction

Michelson interferometers have been developed by various Gravitational-wave observatories (LIGO) [1], TAMA 300 [2], VIRGO [3], GEO 600 [4] etc., where an incident gravitational wave is expected to displace the end mirrors in the two arms by different amounts, and thereby generates a phase shift between the interfering light beams. For high sensitivity the arms of the interferometer must be long including multiple back-and-forth reflections in each arm. Also, an intense laser beam is needed, so that even a quite very small phase shift is expected to be detectable

It has been claimed that in order to achieve maximum sensitivity, it is desirable to adjust the interferometer so that in the absence of the gravitational wave the light beams interfere destructively. When a gravitational wave changes the phase between the emerging light beams a light signal will appear in the "dark output port" of the interferometer. While this conclusion is correct and based on well established analysis for measuring interferometric light intensity, the conclusion is quite different if the analysis is related to second order correlation function measurements [5], as is described in the present analysis. According to the conventional approach the measurement must be completed within half a cycle of the gravitational wave, since after half a cycle the wave will displace the end mirrors in opposite directions, and the phase shift between the interfering light beam will return to zero. It is suggested here to keep the interferometer at a mean phase difference $\phi_0 = \pi/2$ where the transfer from phase fluctuations $\Delta\phi(t)$ to light output intensity fluctuations $I(t) - I_0$ is maximal, where $I_0$ is the average light intensity, giving [6]

$$I(t) = I_0 \left(1 - \Delta\phi(t)\right) \quad . \tag{1}$$

The second order correlation function is given by $\langle I(t)I(t+\tau)\rangle$ where averaging is made on the initial time t, for times much longer than the expected gravitational wave period, and the correlation is a function of the delay time $\tau$. Also to get significant results one has to repeat many times the measurements (in order of thousands or more) for a certain time $\tau$. We show that the measurements of second order correlation functions for the photocurrents or photonumbers can lead to gravitational waves detection. This method is expected to be efficient only for detecting periodic coherent gravitational waves and is based on a quite complicated data processing, but can be made in a straightforward way. The Fourier transform of the correlation function has been used in the analysis of reduction of quantum noise by squeezed states [7,8]. Here it is claimed that such Fourier transform can lead to signals which will appear in the noise background. The Michelson interferometer includes various noise terms [9]. Following conventional analysis [10] the gravitational wave frequencies for detecting them on earth is in the region $10-10^4$ hertz. The amount of gravitational energy on earth, in less than half a cycle, is then extremely small. The advantage in using the second order correlation function and respectively its Fourier transform is that the information on gravitational waves can be obtained from the gravitational wave energy of $10^6$ cycles or more, depending on the coherence time of the gravitational wave as function of $\tau$.

**2. Reduction of quantum noise by the use of squeezed states in Michelson interferometer.**

In order to avoid seismic noise in Michelson interferometer the mirrors are suspended from wires, like pendulums. If the frequency of the pendulum is much lower than the frequency of the gravitational wave, then the suspended mass will behave like a free mass. The interferometer is also operated in vacuum and by avoiding various noise mechanisms the shot noise and the radiation pressure become the dominant noise terms in the present observatories for gravitational wave detection. Usually the Fourier transform of the correlation function is measured by homodyne detection of the respective currents in the detectors. The mean–squared value of the shot noise current is given by [11]

$$\langle i_{sh}^2 \rangle = 2qiB \quad (A^2) \tag{2}$$

where

$$\begin{aligned} q &= 1.60219 \cdot 10^{-19} \ C \quad (\text{the electronic charge}) \\ i &= \text{average current, in amps} \\ B &= \text{bandwidth, in hertz} \end{aligned} \tag{3}$$

To combine the effects of two or more independent noise sources we must add mean squared voltages or currents from the various sources.

The responsivity $\Re_0$ of a photodiode is defined as the output photocurrent produced per unit of incident optical power where its units are amperes per watt (A/W). An incident optical power $P_i$ with frequency $\nu$ is equivalent to $P_i/h\nu$ photons per second ; h is Planck's constant. Let $\eta$ be defined as the ratio of the average number of emitted electrons to the number of incident photons. The average number of electrons emitted per second will then be $\eta P_i/h\nu$. The photocurrent is

$$i_{ph} = q\eta P_i / h\nu \quad (A), \tag{4}$$

and the responsivity is

$$\Re_0 = \frac{i_{ph}}{P_i} = \frac{q\eta}{h\nu} \quad (A/W). \tag{5}$$

It is sometimes useful to express $\Re_0$ as

$$\Re_0 = \frac{\eta q \lambda}{hc} = \frac{\eta \lambda}{1.24} \quad (\lambda \text{ in } \mu m). \tag{6}$$

Let us assume that the light wave is intensity modulated with modulation index m

$$P_i = P_0 [1 + mf(t)] \quad (W) \tag{7}$$

The photocurrent is

$$i_{ph} = \Re_0 P_0 [1 + mf(t)] \quad (A) \tag{8}$$

The average photocurrent is

$$\langle i_{ph} \rangle = \Re_0 P_0, \tag{9}$$

and the signal component of the photocurrent is

$$i_s = \langle i_{ph} \rangle mf(t) \quad (A). \tag{10}$$

The signal-to-noise ratio is defined as the ratio of signal power to noise power or the ratio of the squared currents. Then the S/N which applies to the photocurrent is given by

$$\frac{S}{N} = \frac{\langle i_{ph} \rangle}{2qB} m^2 \langle f^2(t) \rangle. \tag{11}$$

Let us put some numbers: by assuming $\eta \approx 1$, $\lambda = 0.5 \ \mu m$, we get

$$\Re_0 = \frac{0.5}{1.24} \quad (A/W). \tag{12}$$

For periodic gravitational waves we can assume $f(t) = \cos(\omega t + \theta)$ where $\omega/2\pi$ is the frequency of the gravitational wave to be measured on earth which can be in the range of $10 - 10^4$ cycles/sec [10]. Let us assume that for a strong gravitational wave $\Delta L / L \approx 10^{-21}$ then for 4km length in the interferometer $\Delta L = 8 \cdot 10^{-16} \ cm$. Dividing this by $\lambda = 0.5 \cdot 10^{-4} \ cm$ we get:

$$m \approx 1.6 \cdot 10^{-11}. \tag{13}$$

Assuming $B = 100 \ hertz$ we get $2qB \approx 3 \cdot 10^{-17}$, and

$$\frac{S}{N} \approx \Re_0 P_0 \cdot 10^{-5} \langle f^2(t) \rangle. \tag{14}$$

Using the value of $\Re_0$ of (12) we find that in order to obtain significant values of S/N for gravitational waves in Michelson interferometer we need power intensities of order $10^6$ *watt* which is not practical.

Caves was the first scientist to propose a squeezed input Michelson interferometer in which a squeezed vacuum replaces the vacuum state entering in one port ,and has shown the advantage of using such scheme [12]. The main result of Caves, is that by using a squeezed–state Michelson interferometer one can reach the SQL with reduced laser power. It has been shown, however, [13,14] that by including both radiation pressure and photon-counting (PC) fluctuations in a single unified calculation the quantum noise can be reduced below the SQL. Let us explain this difference in the results as follows:

Caves [12] has used the following transformation for the interferometer's beamsplitters :

$$\hat{b}_1 = 2^{-1/2}(\hat{a}_1 + e^{i\mu}\hat{a}_2) \quad , \quad \hat{b}_2 = 2^{-1/2}(\hat{a}_2 - e^{-i\mu}\hat{a}_1) \quad . \quad (15)$$

Here $\hat{a}_1$ , $\hat{a}_2$ are the annihilation operators of the beamsplitter's two 'in' modes and $\hat{b}_1$ , $\hat{b}_2$ are the operators for the two 'output' modes . $\mu$ is the relative phase shift and in (15) the overall phase shift is ignored as it does not affect the results. The coherent state $|\alpha\rangle$ has a phase dependence $e^{i\psi}$ defined by the relation $\alpha = |\alpha| e^{i\psi}$. The squeezed state is defined by using the parameter $\xi = -re^{i\theta}$, with $r > 0$ [12]. By taking into account the correlations between the radiation pressure and PC noise terms maximal reduction in the quantum noise is obtained under the condition

$$\xi = 2\mu + \theta - 2\psi = 0 \quad (16)$$

and under this condition the quantum noise can be reduced below the SQL.

For the special case of $\xi = \pi$ the PC fluctuations (squared) are reduced by the factor exp(-2r) , while the light-pressure-induced fluctuations (squared) are increased by the factor exp(2r). These results correspond to the results obtained by Caves' analysis [12]. However in his basic studies of the effects of squeezed states he has not taken into account the correlations between the PC and and radiation-pressure fluctuations under which both fluctuations can be decreased by the special choice of the phase $\xi$. Similar conclusion has been obtained also in another article [15]. That both fluctuations can be decreased by the use of squeezed states has a very important implication for gravitational waves detection. The reduction of the quantum noise below the shot noise has been demonstrated in recent articles [16] by the use of squeezed states with advanced interferometric techniques.

**3. The use of second order correlation functions for gravitational waves detection**

Using the relation (10) the second order correlation function for the photocurrent fluctuations is given by

$$G(\tau) = \langle i_{ph} \rangle^2 m^2 \langle f(t) f(t+\tau) \rangle = \frac{1}{2} \langle i_{ph} \rangle^2 m^2 \cos(\omega\tau + \theta) \exp(-\gamma\tau) \ . \quad (17)$$

Here we assumed a periodic gravitational wave with angular frequency $\omega$ (with a constant phase $\theta$) and where the correlation time $\frac{1}{\gamma}$ can be few hours (i.e., of times for which the location of the interferometer on earth is not changed significantly relative to the gravitational wave). One should take also in account that for strong gravitational waves which are quasi-periodic the correlation time will be much smaller.

By using a spectrum analyzer, as was applied in experiments on quantum noise reduction [7,8], we will get the spectrum of the gravitational wave

$$G(\omega') = \int G(\tau) \exp(-i\omega'\tau) d\tau \approx \frac{1}{2} \langle i_{ph} \rangle^2 m^2 \frac{\gamma}{(\omega'-\omega)^2 + \gamma^2} \quad , \quad (18)$$

which is expected to appear as a very narrow peak in the spectrum analyzer measurements.

Let us put here again some numbers: The shot noise appears approximately as a constant noise in the frequencies' spectrum of the spectrum analyzer and is given by

$$G_{sh} = 2qi_{ph} \approx 3.2 \cdot 10^{-19} \langle i_{ph} \rangle \qquad . \tag{19}$$

This noise result from the discrete property of the photons and has also a quantum mechanical explanation [5]. We find that the maximal ratio between the gravitational wave signal $G(\omega')$ of (18) and the shot noise $G_{sh}$ of (19) is given by

$$\left( \frac{G(\omega')}{G_s} \right)_{max} = \frac{1}{4} \langle i_{ph} \rangle \frac{m^2}{q\gamma} \tag{20}$$

Estimating the value of m according to (13) and using (9) the ratio (20) is transformed into

$$\left( \frac{G(\omega')}{G_s} \right)_{max} \approx \frac{1.6}{4} \cdot \frac{10^{-3}}{\gamma} \Re_0 P_0 \quad, \tag{21}$$

where $\Re_0$ is given by (12). We find that if $\frac{1}{\gamma} > 10^3$ (sec) we can observe the periodic gravitational wave by this method.

**4.Discussion**

For periodic gravitational waves the information from second order correlation functions is obtained from times which are much longer than the half cycle measurement time assumed in the conventional method. The second order correlation function is obtained by a quite complicated data processing and the gravitational signals are expected to appear in the frequencies' Fourier spectrum of a spectrum analyzer. The second order correlation function can be measured either in photon currents or in photon numbers.

One should take into account that $\frac{\Delta L}{L}$ for periodic gravitational waves might be of order $10^{-23}$ which is of two orders of magnitude smaller than that assumed in the above calculations. However, by using squeezed states in Michelson interferometer the reduction of the quantum noise can compensate the possibility of weaker intensities of gravitational waves. Measurements of the internal frequencies of the interferometer can be avoided by comparing the measurements of two different interferometers. In conclusion , the detection of gravitational can be obtained by the above complicated data processing.